# Disorder induced multifractal superconductivity in monolayer niobium dichalcogenides

Kun Zhao,[1,6,*] Haicheng Lin,[1,6,*] Xiao Xiao,[2,*] Wantong Huang,[1,6] Wei Yao,[1,6] Mingzhe Yan,[1,6] Ying Xing,[3,6] Qinghua Zhang,[4,6] Zi-Xiang Li,[1,6] Shintaro Hoshino,[5] Jian Wang,[3,6] Shuyun Zhou,[1,6], Lin Gu,[4,6] Mohammad Saeed Bahramy,[5] Hong Yao,[1,6] Naoto Nagaosa,[5] Qi-Kun Xue,[1,6] Kam Tuen Law,[2] Xi Chen,[1,6,†] and Shuai-Hua Ji[1,5,6,†] $NbSe_2$ shows superconductivity in the two-dimensional (2D) limit[20-23]. Monolayer $NbSe_2$ consists of one layer of Nb atoms sandwiched between two layers of Se atoms. The unit cell is half of that for $2H$-$NbSe_2$. The atomic layers have hexagonal lattice and are arranged in a trigonal prismatic structure with no inversion symmetry (Fig. 1(a)). Superconductivity in such a non-centrosymmetric structure provides a platform to study the interplay between spin-orbit coupling and superconductivity. It has been shown that the strong spin-orbit coupling is responsible for the large enhancement of the upper critical field and the creation of unconventional pairing correlations in the superconducting states[23,24]. Besides these striking features, the two-dimensional geometry makes it possible to introduce disorder in a controllable way through either adatom adsorption or isovalent substitution during the sample growth. Thereby the superconductivity properties of $NbSe_2$ monolayers can be systematically studied as a function of disorder strengths. In particular, a regime far beyond the scope of application for the Anderson's theorem[1,25] is reached where superconductivity is strongly enhanced. In this regime, the eigenstate correlation exhibits a power-law behavior[26,27], which is a signature of multifractal wave function. The localization[28,29] induced multifractality of electron wave functions enhances the effective electron-electron interaction and hence the superconductivity[18,19,30].

[1]*State Key Laboratory of Low-Dimensional Quantum Physics, Department of Physics, Tsinghua University, Beijing 100084, China*

[2]*Department of Physics, Hong Kong University of Science and Technology, Clear Water Bay, Hong Kong, China*

[3]*International Center for Quantum Materials, School of Physics, Peking University, Beijing 100871, China*

[4]*Beijing National Laboratory for Condensed Matter Physics, Institute of Physics, Chinese Academy of Sciences, Beijing 100190, China*

[5]*RIKEN Center for Emergent Matter Science (CEMS), Wako 351-0198, Japan*

[6]*Collaborative Innovation Center of Quantum Matter, Beijing, China*

---

\* These authors contribute equally to this work.

† Correspondence should be sent to xc@mail.tsinghua.edu.cn and shji@mail.tsinghua.edu.cn

**The interplay between disorder and superconductivity is a subtle and fascinating phenomenon in quantum many body physics. The conventional superconductors are insensitive to dilute nonmagnetic impurities, known as the Anderson's theorem[1]. Destruction of superconductivity and even superconductor- insulator transitions[2-10] occur in the regime of strong disorder. Hence disorder-enhanced superconductivity is rare and has only been observed in some alloys or granular states[11-17]. Because of the entanglement of various effects, the mechanism of enhancement is still under debate. Here we report well-controlled disorder effect in the recently discovered monolayer $NbSe_2$ superconductor. The superconducting transition temperatures of $NbSe_2$ monolayers are substantially increased by disorder. Realistic theoretical modeling shows that the unusual enhancement possibly arises from the multifractality[18,19] of electron wave functions. This work provides the first experimental evidence of the multifractal superconducting state.**

The experiments were conducted on a dilution-refrigerator based ultra-high vacuum (UHV, $1\times10^{-10}$ torr) scanning tunneling microscope (STM) equipped with molecular beam epitaxy (MBE). At the base temperature, fitting the differential conductance of superconducting gap gives the electron temperature of 228 mK (Supplementary Information). The NbSe$_2$ monolayer was prepared on the graphitized 6*H*-SiC(0001) substrate mainly terminated with epitaxial bilayer graphene. High-purity Nb and Se were co-deposited onto the substrate held at 650 °C. After single layer of NbSe$_2$ was formed, the sample was immediately transferred into the STM head to perform local topographic imaging and scanning tunneling spectroscopy (STS) measurements.

The large-scale topographic image in Fig. 1(b) shows the typical sample surface. The NbSe$_2$ monolayer fully covers the graphene/SiC(0001) substrate and forms large atomically flat terraces. The atomic steps indicated by arrows are copied from the substrate. Similar to the epitaxial graphene on SiC(0001) surface, the lattice of NbSe$_2$ is continuous across the steps (Supplementary Fig. S5(b)). The inset of Fig. 1(b) shows the atomically resolved STM image of monolayer NbSe$_2$ surface. The protrusions correspond to the topmost Se atoms. In addition, the superlattice of charge-density wave (CDW) with a momentum of $q_{CDW} \sim 1/3 k_{Bragg}$ is clearly resolved, which is almost identical to the bulk counterpart[31]. The lattice structure is also confirmed by *ex situ* scanning transmission electron microscopy (STEM) measurement on a sample covered by amorphous Se capping layer. Figure 1(c) shows the high-angle annular dark-field (HAADF) image of NbSe$_2$ monolayer along the $[10\bar{1}0]$ direction. Three atomic layers can be easily distinguished from the nearby material, indicating a sharp interface between epitaxial NbSe$_2$ monolayer and the substrate. The atomic structure revealed by STEM is consistent with the lattice side view in Fig. 1(a), unambiguously confirming the non-centrosymmetric structure of the epitaxial NbSe$_2$ monolayer. The electronic band structure of NbSe$_2$ monolayer is characterized by angle-resolved photoemission spectroscopy (ARPES) measurement, which agrees well with the first principle calculation (Supplementary Information).

Superconductivity of NbSe$_2$ survives in the 2D limit. STS reveals the superconducting gap of pristine NbSe$_2$ monolayer (Fig. 1(d)). The variable-temperature spectra are shown in the inset of Fig. 1(d) and the Supplementary Fig. S6. The small dips outside the coherence peaks could be traced to the phonon modes. Elevated temperature suppresses the superconducting coherence peaks and raises the zero-bias conductance (ZBC). The spectra are well fitted by using the BCS gap function. Compared with the bulk transition temperature of 7.2 K (see ref. 32), $T_c$ of monolayer is reduced to about 0.9 K found by fitting either the superconducting gap (Fig. 1(d)) or the ZBC alternatively (Supplementary Fig. S7) as functions of temperature.

Well-controlled experiments of disorder effect were performed on NbSe$_2$ monolayer under an UHV condition. The STS measurement of superconducting gap reveals a large enhancement of superconductivity by disorder.

In the first case, we introduce disorder *in situ* by depositing silicon atoms on the surface of clean NbSe$_2$ film. The strength of disorder can be easily tuned by the coverage of silicon adatoms. Figure 2(a) shows the typical topography of monolayer NbSe$_2$ surface with randomly distributed silicon atoms. The d$I$/d$V$ spectra of NbSe$_2$ monolayer were measured for various Si adatoms coverage (Fig. 2(b)). The spectra are highly homogeneous even with Si adatoms (Supplementary Figs. S9 and S10). With gradually increasing disorder, the superconducting gap reaches a maximum and then starts to decrease. The dome-shaped superconducting phase regime in Fig. 2(c) shows that the gap at the optimal coverage of Si adatoms (0.53 nm$^{-2}$) is more than three times larger than that for a pristine film. The critical field is also significantly enlarged by disorder (Supplementary Fig. S11).

In the second case, we substitute Se by isovalent S atoms in NbSe$_2$. The darker spots in the STM image (Fig. 2(d)) are the sulfur atoms in the NbSe$_{2-x}$S$_x$ monolayer. The composition is determined by the flux ratio between Se and S molecular beams during sample growth and can be estimated by counting the number of Se and S atoms in an image. More topographic images of NbSe$_{2-x}$S$_x$ monolayer with different composition are shown in Supplementary Fig. S12. As the concentration of sulfur increases, the evolution of d$I$/d$V$ spectra (Fig. 2(e)) exhibits similar behavior as that in the first case. The maximum superconducting gap is reached at x=0.49. At this optimal condition, $T_c$ (determined by the disappearance of coherence peaks) is as high as 2.9 K (Fig. 2(f)). Further increasing of sulfur concentration suppresses superconductivity.

Further characterization of the $NbSe_{2-x}S_x$ monolayer is shown in Figs. S13-S16. The Fermi momentum $k_F$, coherence length $\xi_{GL}$ and mean free path $\ell$ for the optimally doped sample (x=0.49) are estimated to be $5 \times 10^9$ m$^{-1}$, 16 nm and 2.2 nm, respectively. The large Ioffe-Regel parameter $k_F\ell \sim 10$ indicates that the sample is in the weak disorder regime and the difference between the mean field BCS temperature and the Berezinsky-Kosterlitz-Thouless transition temperature is small[33-39]. Therefore it is still safe to characterize the superconducting transition by the mean field BCS temperature.

The above two different ways of introducing disorder have very similar effect on the superconducting transition of $NbSe_2$ monolayer. Conceivably, they should share the same mechanism. We tend to attribute the superconductivity enhancement to the formation of multifractal electron wave functions, which can strengthen the local two-body interactions[30]. It is well established that for disordered system close to the metal-insulator transition the eigenfunctions are multifractal[26], i.e., the moments of probability distribution scale with an infinite hierarchy of exponents[40]. The physical origin of the enhancement of superconductivity by multifractality is twofold[30]: (i) the eigenfunction occupies only part of the available space, which by normalization enhances its amplitude; (ii) eigenfunctions within a thin energy shell overlap strongly.

Although several theoretical works have proposed that localization induced multifractality can enhance superconductivity[18,19], adequate experimental evidence has been absent because of the difficulty to exclude other mechanisms and control disorder effects precisely. In the present work, the well-controlled experimental conditions provide new opportunities to investigate the enhancement of superconductivity through the multifractality mechanism. Moreover, the high density of state at Fermi level of $NbSe_2$ monolayer and the double-layer graphene substrate screen the long-range Coulomb interaction, which otherwise tends to reduce the effect of multifractality[19]. Here it should be noted that the samples in the study are still in the weak localization regime as mentioned above. Nevertheless, previous simulation and theoretical analysis show that even away from the critical point of Anderson transition, the eigenfunction statistics still exhibit the remnant of multifractal characteristic[19,

[30, 41].

To understand the observed enhancement of superconductivity by disorder, we performed self-consistent mean field calculations. The details of the Hamiltonian are presented in the Supplementary Information. The superconducting gap equation[25]

$$\tilde{\Delta}_m = g \sum_{|\varepsilon_n|<E_D} \frac{M_{m,n}\tilde{\Delta}_n}{\sqrt{\varepsilon_n^2 + \tilde{\Delta}_n^2}},$$

is solved with realistic parameters and disorder strengths, where $\tilde{\Delta}_n$ and $\varepsilon_n$ are the superconducting gap and eigenenergy for state $n$, respectively. $E_D$ is the Debye energy, $g$ is the strength of effective attractive interaction, and $M_{m,n} = \int d^2 r \psi_{m,\uparrow}^*(\vec{r})\psi_{m,\downarrow}^*(\vec{r})\psi_{n,\downarrow}(\vec{r})\psi_{n,\uparrow}(\vec{r})$ is the correlation function of two eigenstates $m$ and $n$ (↑ and ↓ denote the spin-up and spin-down components for a particular state). The self-consistent calculation of gap equation is summarized in Fig. 3(a). Indeed, disorder first enhances superconductivity before suppressing it. Qualitatively, the gap equation suggests two ways to tune superconductivity by disorder: either through $M_{m,n}$ or through the density of states within the Debye energy window. In the regime where Anderson's theorem applies, both $M_{m,n}$ and density of states are not affected by the weak disorder (Supplementary Fig. S18). Further increasing of disorder strength causes $M_{m,n}$ to grow dramatically due to the multifractality of wave function. Meanwhile, the number of states within the Debye energy window may decrease, but at a rate that cannot compensate the effect of multifractality. As a result, superconductivity can be enhanced near the optimal disorder strength. Even stronger disorder drives the system to a regime where the mean energy-level spacing increases very quickly and greatly reduces the density of states within the Debye energy window resulting in the suppression of superconductivity.

To further elucidate the mechanism of superconductivity enhancement from the self-consistent gap equation, we calculated $M_{m,n}$ as a function of energy difference between two eigenstates as shown in Fig. 3(b). In the regime where superconductivity is enhanced, $M_{m,n}$ increases with the disorder strength and shows a power-law dependence on the energy difference between $\varepsilon_m$ and $\varepsilon_n$, namely $M_{m,n} \propto |\varepsilon_m - \varepsilon_n|^{-\gamma}$ and $\gamma > 0$. This power-law behavior signifies the wave function multifractality[26, 27]. We note that the scattering lengths in

the multifractality regime are about a few nanometers and are compatible with those measured by experiments (Supplementary Information). This calculation strongly suggests that the enhancement of superconductivity is a result of the multifractal wave functions of the monolayer NbSe$_2$.

So far the theoretical modeling based on wave function multifractality has quantitatively explained the superconductivity enhancement in monolayer NbSe$_2$. In principle, the direct observation of multifractality requires probing the spatial distribution of individual quantum states with atomic resolution, which is beyond the capability of STM (only measures local density of states (LDOS)). However, the usefulness of STM to probe multifractal wave-function is based on the following property of multifractality: eigenfunctions within a thin energy shell overlap strongly[30]. Therefore multifractality can be readily revealed by the inhomogeneity[42-44] in LDOS. Figures 4(a) and (b) present the spectra between -2 mV and 2 mV for pristine NbSe$_2$ and NbSe$_{2-x}$S$_x$ monolayer films, respectively. The spatial fluctuation of LDOS is mainly manifested in the coherence peaks. In contrast, the amplitude of order parameter (see the statistics of gap size in Fig. S19) is more robust against disorder in the weak localization regime (for example, Ref. 45). The observed inhomogeneity is consistent with the theoretical modeling (Fig. 4(c)). The spatial fluctuation of LDOS is much more evident in the tunneling spectra from -20 to 20 meV (Figs. 4(d) and (e)) with a variation up to 30% in the doped sample. In addition, the LDOS is highly asymmetric with respect to the Fermi level as predicted by theory[46].

The spatial fluctuation of LDOS is also evident in the dI/dV mapping at various bias voltages (Figs. S20-S21). Multifractality in the mapping of LDOS is quantitatively characterized by the self-similarity at different length scales through the singularity spectrum $f(\alpha)$[42,47]. The deviation of the maximum of $f(\alpha)$ (Fig. S22) from 2 indicates the anomalous scaling in two dimensions.

To identify wave function multifractality more definitively as the major reason for superconductivity enhancement, contributions from other mechanisms, such as carrier doping, suppression of CDW, *etc*., have to be carefully analyzed.

Carrier doping is negligible and does not have prominent contribution to the enhancement of superconductivity. All the major features in the d$I$/d$V$ spectra from -1.0 V to 0.5 V (Supplementary Fig. S8), such as peaks and shoulders, are almost identical for both pristine and Si-doped NbSe$_2$ monolayer. No notable charge transfer has happened between Si adatoms and NbSe$_2$ film. A similar situation shows up in the NbSe$_{2-x}$S$_x$ monolayer, where S isovalently substitutes Se and does not introduce carrier to the electronic band derived from the d orbitals of Nb at the Fermi level.

Superconductivity and CDW coexist in NbSe$_2$ monolayer. In some materials (for example, refs. 48, 49), superconductivity and CDW are competing orders with similar energy scales and the suppression of CDW could lead to the enhancement of superconductivity. However, superconductivity and CDW are weakly linked in NbSe$_2$. In the bulk form, both of them are simultaneously suppressed by substituting Se with S atoms[50]. As a matter of fact, only 1% of the total density of states at the Fermi surface of bulk NbSe$_2$ is involved in the formation of CDW[51-53]. Bulk NbSe$_2$ is a layered material and we expect the NbSe$_2$ monolayer to have similar level of correlation between superconductivity and CDW.

Other possible mechanisms may involve chemical pressure and elevated Debye temperature. Substitution of Se by smaller S atoms shrinks the lattice (Supplementary Fig. S14) and the lighter S atoms can lead to higher Debye temperature. However, these two effects do not exist in the case where similar superconductivity enhancement is also induced by Si adatoms and so can be excluded in general. In addition, previous experimental evidences[54] indicate that a pressure equivalent to the lattice compression in the order of 0.1 Å only results in an insignificant increasing of $T_c$. By putting together all the above considerations, we propose that the enhancement of superconductivity in 2D NbSe$_2$ monolayer is mainly due to the wave function multifractality. Extensions to other 2D or strongly correlated superconductors need further investigations.

**Methods**

The bilayer graphene[55] was produced on $n$-type 6$H$-SiC(0001) (resistivity: 0.02-0.2 Ω·cm) surface by thermal desorption of Si. The growth of NbSe$_2$ was under a Se-rich condition and monitored by *in situ* reflection high-energy electron diffraction (RHEED). The growth rate was about 2.5 monolayer per hour. A polycrystalline Pt-Ir alloy tip was used for STM measurements. Tip was further modified and calibrated on a clean Ag(111) surface. The d$I$/d$V$ spectra were acquired by the standard lock-in technique. STM images were processed using WSxM software[56]. For subsequent *ex-situ* STEM and ARPES measurements, amorphous Se capping layer of about 20 nm thick was deposited on the NbSe$_2$ film at 80 K in UHV to avoid contamination and oxidation in the ambient condition. For ARPES measurements, the Se capping layer was removed by annealing the sample at 300 ℃ for about 20 minutes.

**Acknowledgements**

We thank A. M. García-García for stimulating discussion. This work is supported by the National Natural Science Foundation of China (Grant No. 51561145005, 11622433, 11574175, 51522212, 11774008, 11704414), Ministry of Science and Technology of China (Grant No. 2013CB934600, 2016YFA0301002, 2017YFA0303302, 2018YFA0305603). KTL would like to acknowledge the support of HKRGC (Grants 16324216, HKUST3/CRF/13G, C6026-16W), Croucher Foundation and Dr. Tai-Chin Lo Foundation. L.G. is partially supported by Strategic Priority Research Program of the Chinese Academy of Sciences (grant No. XDB07030200). M.S.B. gratefully acknowledges support from the CREST, JST (No. JPMJCR16F1) and the Japan Society for Promotion of Science (Grant-in-Aid for Scientific Research (S) No. 24224009). During the submission, we also noticed similar results have been reported in the reference arXiv:1810.08222.

**Author contributions**

S.H.J. and X.C. coordinated the project and designed the experiments. K.Z., H.C.L., W.T.H. performed the MBE growth and STM experiments. X.X. and K.T.L. interpreted the results through model calculation. W.Y., M.Z.Y., S.Y.Z. contributed the ARPES measurement. Q.H.Z. and L.G. contributed the STEM characterization. M.S.B. calculated the electronic band by density function theory. Z.X.L., S.H., H.Y, N.N. contributed part of theoretical analysis. K.Z., H.C.L., X.X., K.T.L., S.H.J. and X.C. co-wrote the paper. All authors discussed the results and commented on the manuscript.

**Figure Captions**

**Figure 1, Epitaxial NbSe$_2$ monolayer. a,** The top and side views of the atomic structure of NbSe$_2$ monolayer. **b,** Topographic image of the epitaxial NbSe$_2$ monolayer on graphene/SiC. Image size 250 nm × 250 nm, sample bias 3.0 V, tunneling current 20 pA, temperature 5.6 K. The inset is the atomic resolved STM image (20 nm × 20 nm, 50 mV, 100 pA, 0.54 K). The trigonal islands are bilayer or trilayer NbSe$_2$. **c,** HAADF image of the cross-sectional NbSe$_2$

monolayer. The cutting direction of the focused ion beam milling is along [01$\bar{1}$0] of NbSe$_2$ monolayer. The purple and yellow spheres indicate the Nb and Se atoms, respectively. **d,** The superconducting order parameter as a function of temperature. The open circles show the measured gaps. The dashed curve is the fitting by using $\varDelta \sim (1-(T/T_c))^{1/2}$, which is valid near the critical temperature. The inset shows the d$I$/d$V$ spectrum on NbSe$_2$ monolayer at base temperature (set point: 1.0 mV, 100 pA; lock-in oscillation amplitude 10 μV).

**Figure 2, Disorder-enhanced superconductivity. a,** The topographic image of NbSe$_2$ monolayer with randomly distributed silicon adatoms (14 nm × 14 nm, 1.0 V, 100 pA). The bright protrusions are silicon adatoms. **b,** The d$I$/d$V$ spectra (set point: 1 mV, 100 pA; base temperature) near the Fermi level for various Si adatom concentration. **c,** The superconducting gap as a function of the Si adatom density. The dashed curve is a guide to the eye. **d,** Atomic resolved image of NbSe$_{2-x}$S$_x$ (x=0.31) monolayer (10 nm × 10 nm, 100 mV, 100 pA). The sulfur atoms are darker in the image. **e,** The d$I$/d$V$ spectra (base temperature) near the Fermi level with various sulfur concentration in NbSe$_{2-x}$S$_x$ monolayer. **f,** The superconducting transition temperature as a function of x. The open squares are the gap closing temperature while the red dots are the temperatures for the disappearance of coherence peaks. The dashed curves are a guide to the eye. More information on analyzing the tunneling spectra is provided in the supplementary materials.

**Figure 3, Theoretical modeling. a,** The superconducting gap as a function of disorder strength characterized by the scattering length. The gap is obtained by solving the self-consistent gap equation and averaging over 35 different disorder configurations. At the optimal disorder strength, the scattering length is $\ell \sim 2.38$ nm, where the superconducting gap is doubled. In experiment, a similar scale ($\ell \sim 2.2$ nm) is extracted from the vortex core (Supplementary Information) at the maximal $T_c$ with superconducting gap enlarged by a factor of three. **b,** The eigenstate correlation function versus the energy difference between the eigenstates for four typical disorder strengths highlighted in **a**. The correlation function shows power-law dependence on the energy difference, which is a defining signature of wave function multifractality.

**Figure 4, Spatially resolved spectra near Fermi level. a,b,** The spatially resolved d$I$/d$V$ spectra across the surface of NbSe$_{2-x}$S$_x$ monolayer (x=0 for **a**, x=0.49 for **b**). The spectra are taken along a line of 26.3 nm for **a** and 55.8 nm for **b**. Set point: 2 mV, 100 pA; lock-in oscillation amplitude: 20 μV; base temperature. **c,** Calculated spatially resolved d$I$/d$V$ spectra with the point space of 4 (in the unit of lattice constant of theoretical model) by using a 4×4 lead, which simulates the finite size of a realistic STM tip. The scanning is along a column of lattice on which the effective tight-binding Hamiltonian is built. Thermal broadening η=0.001t. **d,** The spatially resolved d$I$/d$V$ spectra of the sulfur doped monolayer NbSe$_2$ with x=0.42. Set point: 100 mV, 100 pA; lock-in oscillation amplitude: 1 mV; base temperature. **e,** The spatially resolved d$I$/d$V$ of the pristine monolayer NbSe$_2$. Set point: 100 mV, 100 pA; lock-in oscillation amplitude: 1 mV; base temperature.

**Figure 1**

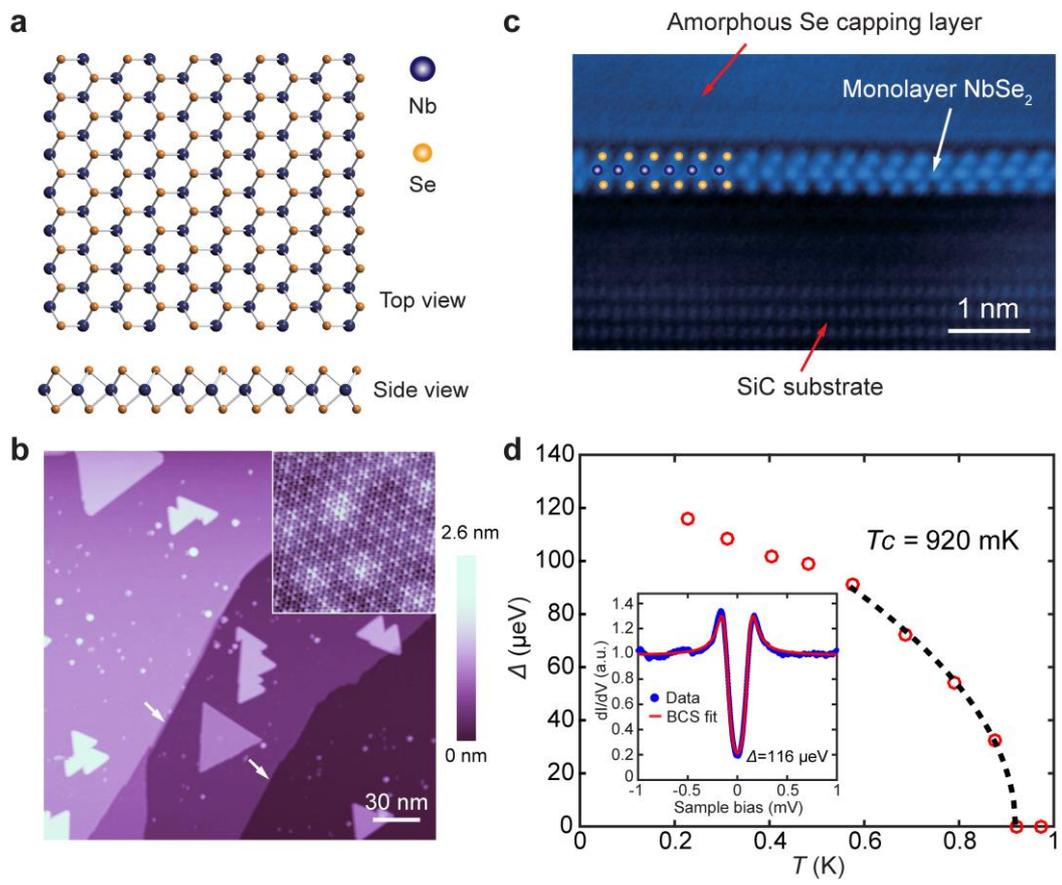

**Figure 2**

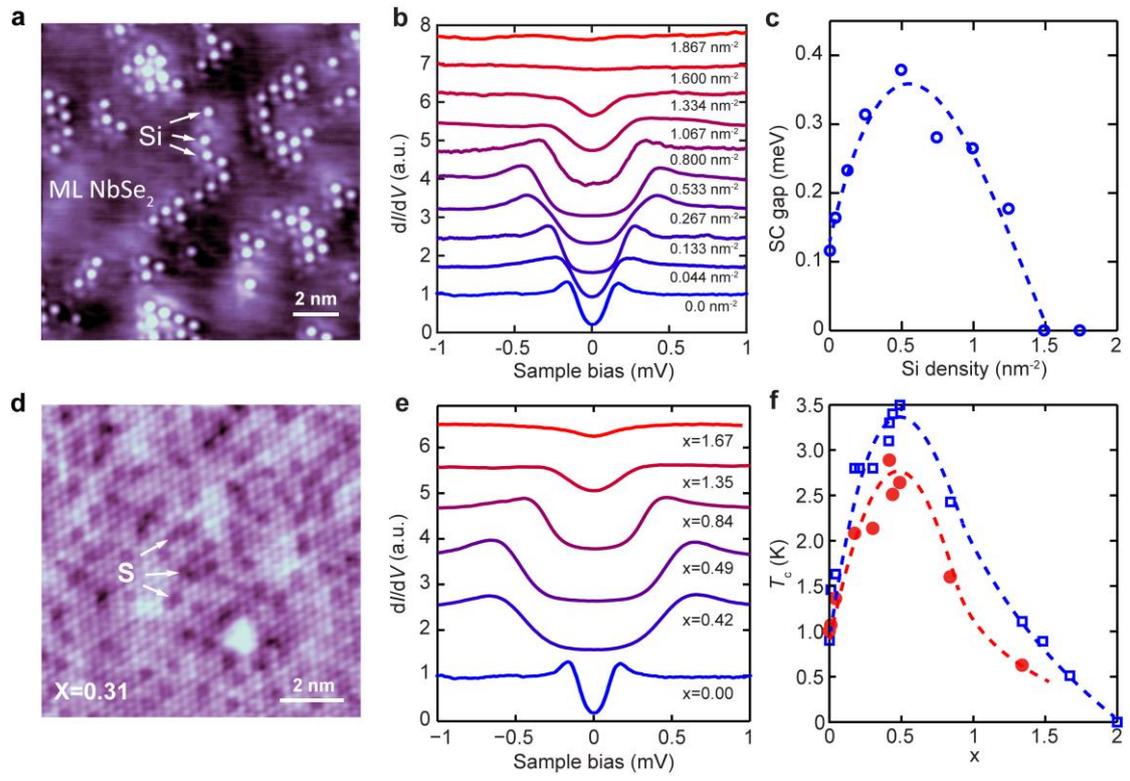

**Figure 3**

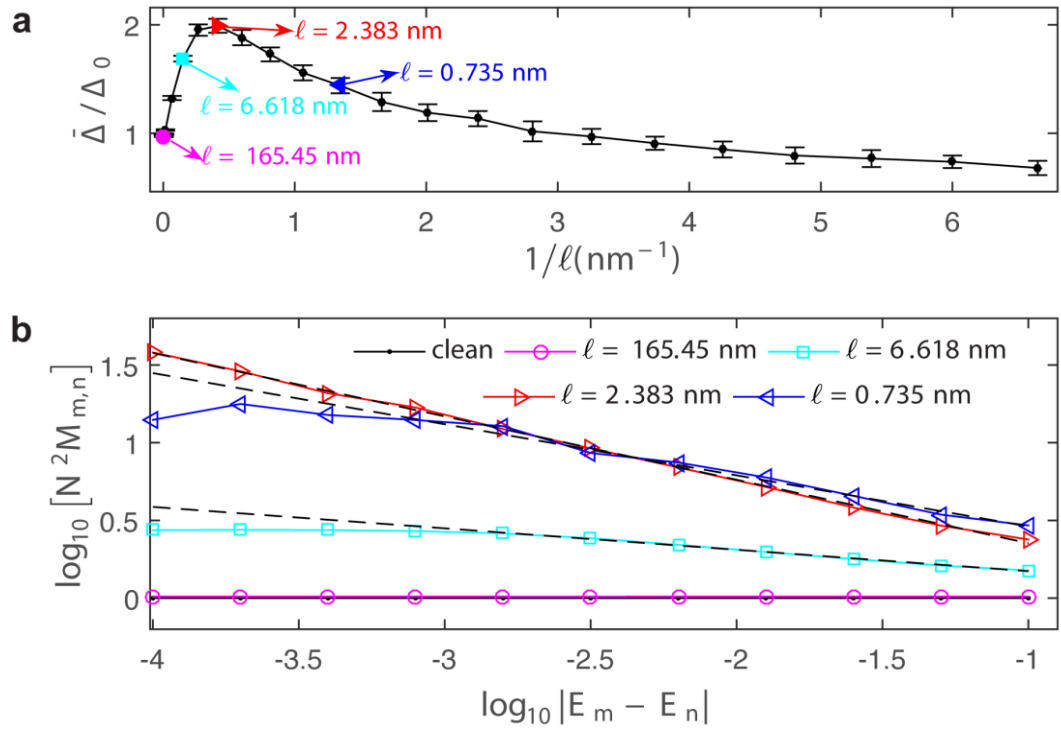

**Figure 4**

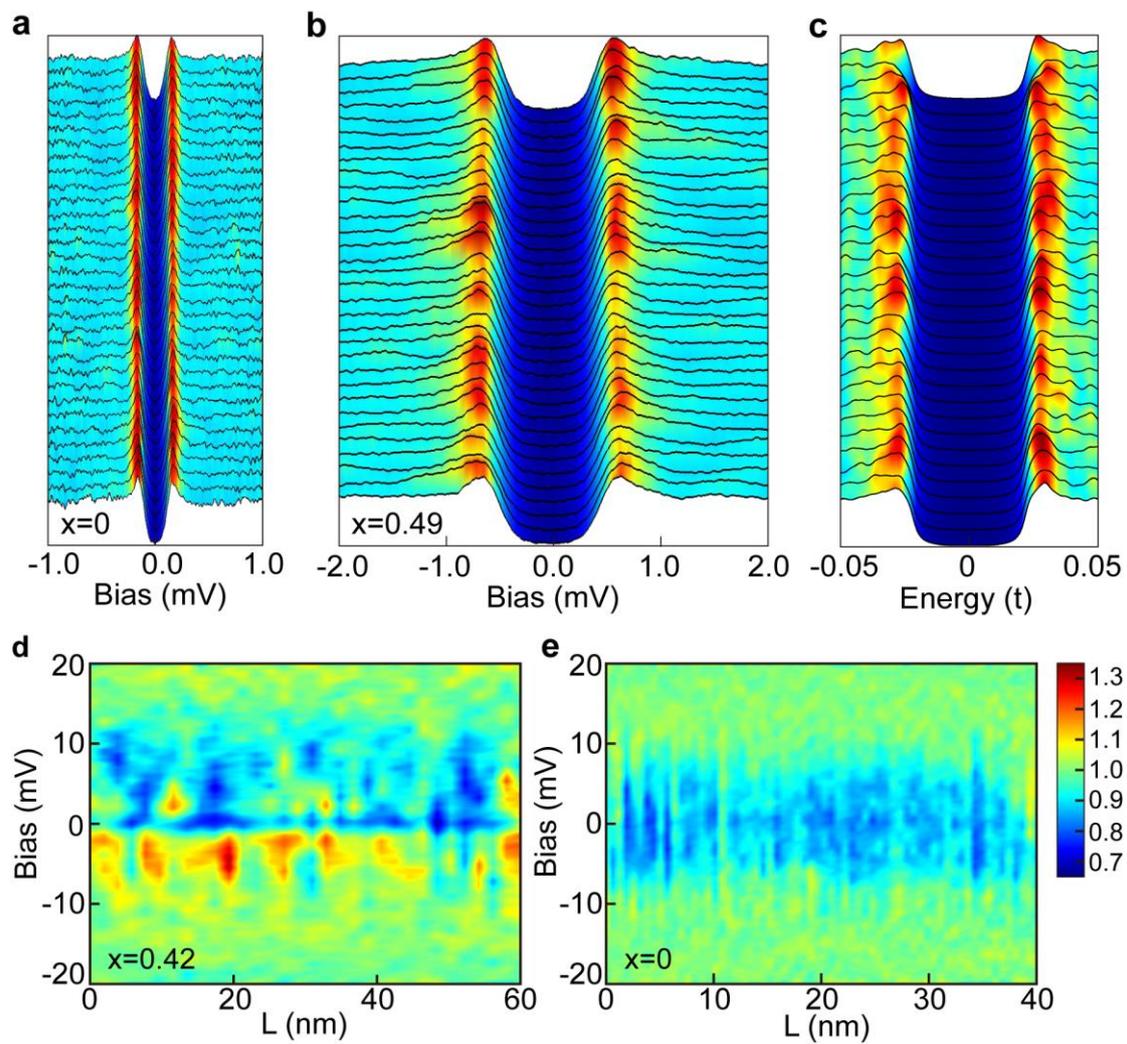